\documentstyle[preprint,aps]{revtex}
\begin{document}
\draft
\preprint{INPP-UVA}
\title{A Phenomenological Theory of Fermion Masses and Mixings}
\author{P. Q. Hung}
\address{Department of Physics, University of Virginia,
Charlottesville, VA 22901}
\date{\today}
\maketitle
\begin{abstract}
A phenomenological theory of fermion masses and mixings is constructed
within the framework of a four- family symmetry. It is found that the most
favored set of relevant CKM elements are $|V_{us}|\approx 0.222$,
$|V_{cb}|\approx 0.044$, $|V_{ub}/V_{cb}|\approx 0.082$,
$|V_{ud}|\approx 0.974$, $|V_{cs}|\approx 0.9736$,
$|V_{cd}|\approx 0.224$ with $\hat{B}_K \approx 0.8$. The top
quark mass is predicted to be 258 GeV at 1 GeV with its physical
mass approximately equal to 153 GeV. The Majorana scale associated
with the fourth neutrino is bound from above to be 6.4 TeV.
\end{abstract}

Despite the impressive agreement of the standard model with experiment,
it is clear that questions such as the origin
of fermion masses and mixing angles cannot be answered solely within
the framework of the standard model.
At the present time, one
does not know at what energy scale ( called Family scale here)
lies the solution to the mass problem.
There is a some belief that such a scale might be very near the Planck
mass where it is hoped that all interactions (including perhaps gravity)
are unified. This might be the case. On the other hand, it is also
possible that the Family scale might not be too much higher than
the electroweak scale.
We shall explore this possibility below.

It is a fact that not only do we have inter-generation mass
splitting but we also have intra-generation mass splitting.
This last splitting breaks explicitely the so-called "custodial"
$SU(2)$ symmetry (in the simplest version of the standard model)
which guarantees the $\rho$ parameter to be equal to unity at
tree level. What is the relationship between the
"custodial" breaking and the behaviour
of the fermion mass matrices?
One first
notices that, if the mass matrices for the
Up and Down sectors were identical (in form and magnitude), the
CKM matrix \cite {cabibbo} would become the unit
matrix and all mixing angles would
vanish. The fact that the CKM mixing angles are non-vanishing
and small is an indication that the fermionic "custodial"
breaking should be non-vanishing and small as well.
By "small" we mean that the contribution to
the $\rho$-parameter is small.
We shall use this hint in the construction of our model.

To address the mass problem,
it is obvious that one has to go beyond the minimal (three
generations) standard model. We
propose in this note a model with four generations and a
family symmetry \cite {hung1}. This (broken) family symmetry gives rise
to mixing terms in the fermion mass matrices at one-loop level.
One of the motivations for having a fourth family here is the
fact that, in our model, the main contribution to the
off-diagonal elements of the mass matrices come from one-loop
diagrams with a scalar exchange, and large Yukawa couplings
are preferable to guarantee that these elements are non-negligible.
The nature of fermion masses and mixing
in our framework is linked to the nature of the scalar
vacuum expectation values and resulting scalar mixing.
We shall
show that the mixing terms in the quark and lepton
mass matrices are intrinsically related. As a consequence,
we shall also show how a rare decay process such as
$K_{L}\rightarrow \mu e$ can set an upper bound on the
Majorana mass scale.

To set the stage for the construction of the mass matrices, we
now list our assumptions. They are:

1) There is a vector-like $SU(4)$ family (gauge) symmetry.

2) The combined  generation-electroweak symmetry is
$SU(4)\otimes SU(2)_{L}\otimes SU(2)_{R}\otimes U(1)_{B-L}$.
The existence of $SU(2)_R$ in our context is linked to the
assumption that the family symmetry is vector-like.

3) There is a Higgs field, $\Phi= T_{i}\Phi^{i}$ with $i=1,\ldots,15$,
which transforms as  $(15,2,2,0)$ and another one, $\phi$,
which transforms as $(1,2,2,0)$ under
$SU(4)\otimes SU(2)_{L}\otimes SU(2)_{R}\otimes U(1)_{B-L}$.
(We are ignoring here scalars which are electroweak singlets.)

4) The left and right-handed quarks transform respectively
as $(4,2,1,1/3)$ and $(4,1,2,1/3)$ while the left and
right-handed leptons transform as $(4,2,1,-1)$ and
$(4,1,2,-1)$. The charged generator is
$Q=T_{3L}+T_{3R}+(B-L)/2$.

5) There exists a Yukawa coupling of the form
\begin{eqnarray}
{\cal L}_{Y} = && {\bar{q}}_{L}( h_{1} \Phi + h_{2} \tilde{\Phi}
+ h_{1}^{\prime}\phi + h_{2}^{\prime}\tilde{\phi})q_{R} + h.c.
\nonumber\\
&&+ {\bar{l}}_{L}(l_{1} \Phi + l_{2} \tilde{\Phi}
+l_{1}^{\prime}\phi + l_{2}^{\prime}\tilde{\phi})l_{R} + h.c.
\end{eqnarray}
where the notations used for quark and lepton fields are
self-explanatory and where
$(\tilde{\Phi}, \tilde{\phi}) \equiv (T_i\tau_2\Phi^{i\star}\tau_2,
\tau_2\phi^{\star}\tau_2)$.

6) The Higgs fields have vacuum expectation values (VEV) \cite {hung1}
\begin{mathletters}
\begin{equation}
\langle \Phi \rangle = ({\langle \Phi_{3} \rangle}/\sqrt{2})
diag(1, -1, 0, 0) +
({\langle \Phi_{8} \rangle}/\sqrt{6})diag(1, 1, -2, 0)
+ ({\langle \Phi_{15} \rangle}/\sqrt{12})diag(1, 1, 1, -3),
\end{equation}
\begin{equation}
\langle \varphi \rangle = diag(\kappa, \kappa^{'}),
\end{equation}
\end{mathletters}
where
\begin{equation}
{\langle \Phi_{3,8,15} \rangle} = diag(v_{\{3,8,15\}}\,, v_{\{3,8,15\}}^{'}),
\end{equation}
The various VEV's will in general be complex and can be parametrized
generically as $v e^{i\delta}$.

7) The breaking of the family symmetry leads to the following mass
matrices for the scalars which connect the fourth family to the
other three and to itself, and which are $\Phi_{1,4}=(\Phi_{9}\mp i \Phi_
{10})/\sqrt{2}$, $\Phi_{2,4}=(\Phi_{11}\mp i \Phi_{12})/\sqrt{2}$,
$\Phi_{3,4}=(\Phi_{13}\mp i \Phi_{14})/\sqrt{2}$, and $\Phi_{44}=\Phi_{15}$,
\begin{equation}
M^2 \left(
\begin{array}{cccc}
1 & a & 0 & 0 \\
a & 1 & b & 0 \\
0 & b & 1 & c \\
0 & 0 & c & 1
\end{array}
\right)\ .
\end{equation}
for the neutral sector, $\Phi^{0}$, and
\begin{equation}
M^2 \left(
\begin{array}{cccc}
1 & a^{\prime} & 0 & 0 \\
a^{\prime} & 1 & b^{\prime} & 0 \\
0 & b^{\prime} & 1 & c^{\prime} \\
0 & 0 & c^{\prime} & 1
\end{array}
\right)\ .
\end{equation}
for the charged sector, $\Phi^{\pm}$. The diagonal elements
are (14), (24), (34) and (44) from top to bottom.
The six coefficients $a,\ldots,c^{\prime}$ will
determine the amount of mixing among different generations.
It is found that the primed and unprimed coefficients
are not too different from each other, which means that the
amount of scalar "isospin" breaking is small.
We shall assume that the mixing
among other generation-changing scalars, if it exists, is negligible
compared with the above and hence the only dominant off-diagonal
elements of the mass matrices are those generated by the above
scalars.

8) Since $\Phi$ transform as $(2,2)$ under $SU(2)_L \otimes SU(2)_R$,
each component $\Phi^i$ takes the form
\begin{equation}
\Phi^i= \left(
\begin{array}{cc}
\phi_1^{0i} & \phi_2^{+i}  \\
-\phi_1^{-i} & \phi_2^{0i\star} \\
\end{array}
\right)\ .
\end{equation}
In the computation of the one-loop contribution to various
off-diagonal elements, both $\phi_1^i$ and $\phi_2^i$ contribute.
We will assume that they both have mass mixing of the form of
Eqs.(4,5). Furthermore, we will assume that the coefficients
$a,\ldots,c^{\prime}$ of the two sectors are proportional
to each other, with a common proportionality coefficient
denoted by $y$.

Flavor-changing neutral currents can be suppressed provided
the masses of the family gauge bosons and mixing parameters
in both the left-handed and right-handed quark "rotation" matrices
are appropriately chosen. Similar considerations apply to the
scalar sector. It is beyond the scope of this paper to present
such an analysis. We shall assume it can be done and shall
concentrate here on the construction of the mass matrices.

Various mixing parameters and phases in the fermion
mass matrices are explicitely computed in terms of parameters
of the basic family symmetry.
Once the mixing parameters are
fixed, the phases are completely determined.
These same parameters
enter in the mass matrices of both quark and lepton sectors
so that the mixing angles and phases of
the two sectors are found to be related.

The mass mixing among the scalars
are assumed to be such that only the following mixings occur:
$\phi_{44}-\phi_{43},\phi_{43}-\phi_{42},\phi_{42}-\phi_{41}$
giving rise to a particular
type of symmetric mass matrices.
In a way, fermion mixing can be seen as a direct consequence of
scalar mixing. If we recall Eq.(2), the breaking of the
family symmetry is assumed to be such that all of the family
changing scalars have vanishing vacuum expectation values (a
familiar assumption). In this case, there is no direct fermion
mixing at tree level coming from the symmetry breaking.
On the other hand,
a general gauge-invariant scalar potential will yield,
after symmetry breaking,
various mass mixings among the scalars. Couplings of the fermions
to these same scalars will give rise to fermion mixing beyond
the tree level in the mass matrices. The
Yukawa couplings that enter these radiative corrections are
much larger than the gauge couplings and, because of that feature,
we shall ignore the contributions coming from the family
gauge bosons.

The diagonalization of Eqs.(4,5) can be done analytically in
a straightforward manner.
Each of these four scalar mass eigenstates can now
couple the fourth generation to all the other three, with
the strength given by the elements of the eigenvectors. The
trivial details are given elsewhere. The only remark one
would like to make here is the fact that the mass scale
$M^2$ mentioned above do not enter the calculations
given below because the results there (inside the logarithms)
are given in terms of mass ratios.

At the one-loop level, the off-diagonal elements of the
fermion mass matrices are generated by the exchange of the
above four scalars and are finite
In our model the twenty four off-diagonal
elements of the up- and down-symmetric mass matrices
of the quark and lepton sectors can be computed in terms
of the six parameters $(a,\ldots,c')$, the four fourth-
generation masses, and a phase difference (to be
specified below). Let us first concentrate on the
following elements in a typical mass matrix: 4-3, 3-2, 2-1.
For the up sector (for quarks), the
one-loop contributions are given by
\begin{mathletters}
\begin{equation}
M_{43}^U= h_2^2\,e^{i\delta_1}\,|M_U^0|\,(\lambda^{\prime\prime}
x e^{-i\Delta} - \frac{1}{r}) L_0^{\prime\prime} ,
\end{equation}
\begin{equation}
M_{32}^U= h_2^2\,e^{i\delta_1}\,|M_U^0|\,(\lambda
x e^{-i\Delta} - \frac{1}{r}) L_0 ,
\end{equation}
\begin{equation}
M_{21}^U= h_2^2\,e^{i\delta_1}\,|M_U^0|\,(\lambda^{\prime}
x e^{-i\Delta} - \frac{1}{r}) L_0^{\prime} ,
\end{equation}
\end{mathletters}
where $r=h_2/h_1$ and $x=|M_D^0|/|M_U^0|$, the ratio
of the fourth-generation quark masses.
For the down sector, one has
\begin{mathletters}
\begin{equation}
M_{43}^D= h_1^2\,e^{i\delta_2}\,|M_U^0|\,(\lambda^{\prime\prime}
e^{i\Delta} - rx) L_0^{\prime\prime} ,
\end{equation}
\begin{equation}
M_{32}^D= h_1^2\,e^{i\delta_2}\,|M_U^0|\,(\lambda
e^{i\Delta} - rx) L_0 ,
\end{equation}
\begin{equation}
M_{21}^D= h_1^2\,e^{i\delta_2}\,|M_U^0|\,(\lambda^{\prime}
e^{i\Delta} - rx) L_0^{\prime} .
\end{equation}
\end{mathletters}
Here, $L_0^{(\prime,\prime\prime)}= L_2^{0(\prime,\prime
\prime)}-L_1^{0(\prime,\prime\prime)}$. Assumption (8) gives
$L_2^{0(\prime,\prime\prime)}=r^2 L_1^{0(\prime,\prime\prime)}$
(here $y=r^2$) so
that $L_0^{(\prime,\prime\prime)}=(r^2-1) L_1^{0(\prime,\prime\prime)}$.
The $L_0$'s are the one-loop contribution to the off-diagonal
elements coming from neutral scalars. The parameters $\lambda^{(
\prime,\prime\prime)}$ represent the contribution of the charged
scalars relative to the neutral ones. They are defined for the
case when $r\neq1$. Assumption (8) gives
$L_2^{+(\prime,\prime\prime)}=r^2 L_1^{+(\prime,\prime\prime)}$.
One then has $\lambda^{(\prime,\prime\prime)}=(r/(r^2-1))(L_1^{+
(\prime,\prime\prime)}/L_1^{0(\prime,\prime\prime)})$.
Notice that the various $L$'s can be explicitely computed
using the mass eigenstates and eigenvectors of Eqs.(4,5). They are
typically of the form: $(Aln(m_1/m_2)+Bln(m_3/m_4))/16\pi^2$, where
$A$ and $B$ are some combinations of the elements of the eigenvectors.
The parameter $\Delta$ is defined as $\Delta= \delta_1 - \delta_2$ where
$m_U^0 \equiv e^{i\delta_1}|m_U^0|$ and $m_D^0 \equiv e^{i\delta_2}|m_D^0|$.
$\delta_{1,2}$ are functions of the phases of
the complex VEV's.

The parameters $\lambda$'s and $L_0$'s depend entirely on the scalar
sector and that is the reason why they appear in both Eq.(7) and
Eq.(8). This will also be the reason why they also appear in the
lepton mass matrices. There one just has to make the replacements:
$h_{1,2} \rightarrow l_{1,2}$, $r \rightarrow r_l =l_2/l_1$ and
$x \rightarrow x_l = |m_E^0|/|m_N^0|$ in Eqs. (7,8).

Notice that we seem to ignore elements such as $M_{42}$ and
$M_{41}$ which can be computed at the one-loop level and
other off-diagonal elements which arise at higher order. The
reason is simply that they are numerically small compared
with the above elements. Within the precision considered in
this paper, they are found to be not so important. A more
detailed and precise analysis to be carried out in a
subsequent work will include these extra corrections. It is
however important to state here the fact that these extra
terms are perfectly {\em calculable} in terms of known
parameters of the model.

Let us now examine Eqs. (7,8) in the case when the "isospin"
breaking parameter $r = h_2/h_1$ becomes unity, i.e. no
"isospin" breaking. A look at the diagonal masses obtained
from Eqs. (1,2,3) reveals that when $r=1$ or equivalently
$h_2=h_1$, one obtains $m_U^0=m_D^0$, $m_t^0=m_b^0$,
$m_c^0=m_s^0$, and $m_u^0=m_d^0$. Now $m_U^0=m_D^0$ means
that the parameter $x=1$ and that $\delta_1=\delta_2$ giving
$\Delta=0$. Eqs. (8,9) then tell us that $M_{43}^u = M_{43}^d$,
$M_{32}^u = M_{32}^d$, and $M_{21}^u = M_{21}^d$. (One can
make similar statements for the other off-diagonal elements.)
This means the the mass matrices for the up and down sectors
are {\em identical} and so are the matrices $U_U$ and $U_D$
which diagonalize them. This now means that the generalized
CKM matrix $V_{CKM} = U_U^{-1} U_D$ is equal to the unit matrix.
All mixing angles {\em vanish}. This is precisely the point
that we have mentioned above. This feature is true regardless
of how much "isospin" breaking there is in the scalar sector,
i.e. regardless of how much mass splitting there is between
the charged and neutral scalars.
In our model it turns out that the magnitude of the CKM elements
is a result of the interplay between the "isospin" breaking
term of the Yukawa sector, $r$, and that of the scalar sector
as defined by the parameters $\lambda^{(\prime, \prime\prime)}$.

In order to carry out a numerical study of mass eigenvalues
and CKM matrix elements, it is found to be more convenient
to parametrize the off-diagonal elements of the mass matrices
in a slightly different way although in principle one can
just use directly Eqs. (7,8). One just has to vary the
parameters described above in order to fit the generalized
CKM matrix elements and the quark masses. This is a perfectly
well-defined task albeit a very time-consuming one. To narrow
down the range of values, we are guided by the hierarchical
nature of the masses and the sizes of the CKM matrix elements.
We shall parametrize the various M's as: ${\cal M}_{43}^U=
c_U\lambda_U^2 e^{i\delta_U^{\prime\prime}}$,
${\cal M}_{32}^U= r_t \lambda_U e^{i\delta_U}$,
${\cal M}_{21}^U= r_t \lambda_U^3 e^{i\delta_U^{\prime}}$,
for the up sector, and
${\cal M}_{43}^D= c_D\lambda_D^2 e^{i\delta_D^{\prime\prime}}$
${\cal M}_{32}^D= r_b \lambda_D e^{i\delta_D}$,
${\cal M}_{21}^D= r_b \lambda_D^3 e^{i\delta_D^{\prime}}$,
for the down sector. The previous quantities are the elements
of ${\cal M}_U$ and ${\cal M}_D$ where $M_U= |M_U^0| e^{i\delta_1}
{\cal M}_u$ and $M_D= |M_U^0| e^{i\delta_2} {\cal M}_d$. The
parameters $r_t$ and $r_b$ are defined as $r_t=|m_t^0|/|m_U^0|$
and $r_b=|m_b^0|/|m_U^0|$. The above phases are defined as
$tan\delta_U=\lambda x sin\Delta/(1/r-\lambda x cos\Delta)$ and
$tan\delta_D=\lambda sin\Delta/(\lambda cos\Delta - rx)$ and
similarly for the primed and double-primed quantities. The phases
$\delta_{1,2}$ can be absorbed in a redefinition of the up
and down quark fields. The parameters $\lambda_{U,D}^{(\prime,
\prime\prime)}$ are defined in terms of those of the fundamental
theory via Eqs. (7,8). In total the ten parameters $\lambda_U$,
$\lambda_D$, $c_U$, $c_D$, $\delta_{U,D}^{(\prime,\prime\prime)}$
are computed interms of seven parameters $r$,$\lambda^{(\prime,
\prime\prime)}$, $L_0^{\prime}/L_0$, $L_0^{\prime\prime}/L_0$,
and $\Delta$. To complete the picture, one has to
specify the diagonal elements coming from the other three
generations for both up and down quarks. In general they have
arbitrary phases and magnitudes. Although the phases can be
similar to those of the fourth generation, we shall treat them
as free parameters. It turns out that the attractive possibility
of having all phases approximately equal to each other can
actually be realized in our model.

After factoring out $|M_U^0| e^{i\delta_{1,2}}$ in the up and
down mass matrices, the first three diagonal elements can be
written as $r_u e^{i\alpha_1}$, $r_c e^{i\alpha_2}$,
$r_t e^{i\alpha_3}$ for the up sector and $r_d e^{i\beta_1}$,
$r_s e^{i\beta_2}$, $r_b e^{i\beta_3}$ for the down sector,
where $\alpha_1= \delta_u-\delta_1$, $\alpha_2= \delta_c-\delta_1$,
$\alpha_3= \delta_t-\delta_1$, and $\beta_1= \delta_d-\delta_2$,
$\beta_2= \delta_s-\delta_2$, $\beta_3= \delta_b-\delta_2$.
Here $r_i=|m_i^0|/|m_U^0|$. ${\cal M}_U$ and ${\cal M}_D$
which are symmetric matrices can be made real by an appropriate
redefinition of the quark phases provided
$\alpha_1=2(\delta_U^{\prime\prime}+\delta_U^{\prime}-\delta_U)$,
$\alpha_2=-2(\delta_U^{\prime\prime}-\delta_U)$,
$\alpha_3=2\delta_U^{\prime\prime}$,
$\beta_1=2(\delta_D^{\prime\prime}+\delta_D^{\prime}-\delta_D)$,
$\beta_2=-2(\delta_D^{\prime\prime}-\delta_D)$,
$\beta_3=2\delta_D^{\prime\prime}$. Assuming that $\delta_u,
\delta_c$, etc... satisfy the previous "reality" condidtions,
it will be seen that the only CP phases which enter the generalized
CKM matrix are $\tilde{\Delta}=\delta_U-\delta_D$,
$\tilde{\Delta}^{\prime}=\delta_U^{\prime}-\delta_D^{\prime}$,
and $\tilde{\Delta}^{\prime\prime}=\delta_U^{\prime
\prime}-\delta_D^{\prime\prime}$. Phenomenologically,
$\delta_U,\delta_D$, etc... are found to be either close to
(but not equal to) $0^{\circ}$ or $180^{\circ}$ so that
$\delta_u,\delta_d$, etc... are almost equal to $\delta_1$
and $\delta_2$. Such a possibility is rather attractive:
all quark masses have roughly similar phases.

To bring ${\cal M}_U$ and ${\cal M}_D$ to a real form, we
redefine the left-handed quark phases by a diagonal matrix
of the form $Q_U=diag(1,e^{-i\phi_1^U},e^{-i\phi_2^U},
e^{-i\phi_3^U})$ for the up sector and
$Q_D=diag(1,e^{-i\phi_1^D},e^{-i\phi_2^D},e^{-i\phi_3^D})$
for the down sector, where
$\phi_1=2\delta^{\prime\prime}+\delta^{\prime}-2\delta$,
$\phi_2=\delta^{\prime}-\delta$,
$\phi_3=\delta^{\prime\prime}+\delta^{\prime}-\delta$
with the appropriate subscripts for up and down.
(There is also a redefinition of the right-handed quark
phases but it is irrelevant for $V_{CKM}$ and we shall
not discuss it here.) The eigenvalues of the real, symmetric
4x4 matrices denoted by ${\cal M}_U^R$ and ${\cal M}_D^R$
can be computed numerically. These eigenvalues are in turn
used to construct the eigenvectors which are then used
to obtain the matrices $R_U$ and $R_D$ which diagonalize
${\cal M}_U^R$ and ${\cal M}_D^R$ respectively. The
generalized CKM matrix is then defined as
$V_{CKM}=Q_U^{-1}R_U^{-1}R_D Q_D$. A full account of our
analysis will be given in a separate publication. We shall
illustrate our model with a few examples in this letter.
A few typical elements are
$V_{us}=\{\lambda_D^3/|\tilde{r}_s| - (\tilde{r}_u/\lambda_U^3)
e^{i(2\tilde{\Delta}^{\prime\prime}+\tilde{\Delta}^{\prime}-2
\tilde{\Delta})} + O(\lambda^3)\}/N_{us}$,
$V_{ub}=(\tilde{r}_u/\lambda_U^2)\{K_u^{-1}e^{i(\tilde{\Delta}^
{\prime}-\tilde{\Delta})}+(K_b/\lambda_U\lambda_D)
e^{i(2\tilde{\Delta}^{\prime\prime}+\tilde{\Delta}^{\prime}-2
\tilde{\Delta})} + O(\lambda^2)\}/N_{ub}$,
$V_{cb}=\{-(K_b/\lambda_D)
e^{i(2\tilde{\Delta}^{\prime\prime}+\tilde{\Delta}^{\prime}-2
\tilde{\Delta})} - (\lambda_U/K_c)e^{i(\tilde{\Delta}^
{\prime}-\tilde{\Delta})} + O(\lambda^5)\}/N_{cb}$.
All other elements of the generalized CKM matrix are easily
computable and they will be given in an extensive version
of this manuscript.
The various $N$'s are normalization factors.
The $\tilde{r}$'s are related to
the ratios of the eigenvalues and $r_t$ or $r_b$. The quantities
$K_i$'s are $K_i=1-\tilde{r}_i-(c_{U,D}\lambda_{U,D}^2)^2/(r_{t,b}
(1-r_i))$ with $i=u,c,b$. One can immediately notice that
the mixing of the "light" generations with the fourth generation
creeps into $V_{ub}$ and $V_{cb}$ via various $K_i$'s which
contain $c_U$ or $c_D$ which enter ${\cal M}_{43}$. This is
because of the way the fourth generation mixes
with the third one in our mass matrices.
The CKM elements which do not
involve the third generation-we are mainly concerned with the
3x3 sub-matrix here-are relatively insensitive to the presence
of the fourth generation. In this sense, the physics of the
third generation (in particular B-physics) indirectly probes
the existence or non-existence of a fourth generation.
Recall that the fourth generation indirectly manifests itself
through the parameter $x$ which enters in all elements of the mass
matrices. It is however the resulting (dominant) mass mixing with
the third generation that manifests itself most visibly in the
CKM elements involving the third generation.

An extensive numerical analysis is underway. We shall give here
some preliminary results. The inputs are given in the form of
$M(r_1,r_2,r_3,1)$ for the masses. We have \cite {hung2}
$|M_U^0|=1 TeV$, $x=
|M_D^0|/|M_U^0|=0.98$, $1 TeV(-.8\times10^{-5}, .00213,.54)$ for
the up sector and $0.98 TeV(.596\times10^{-5},-0.952\times10^{-5},
.0062)$ for the down sector, $\lambda_U=0.064$, $\lambda_D=0.15$,
$c_U=112.5$, $c_D=0.325$, $r=1.4637$. The previous values not only
determine the mass eigenvalues but they also fix the phases in the
following way. By equating the two ways of writing $M_{43}$, $M_{32}$,
and $M_{21}$ for the up and down sectors and by taking the absolute
value of these elements, we can fix the values of $\lambda^{
\prime}$ and $\lambda^{\prime\prime}$ once $\lambda$ and
$\Delta$ are given. Let us recall that the phases $\delta_U$,
$\delta_D$ and their primed and double-primed counterparts
are given in terms of $x$, $r$ and $\lambda$, $\lambda^{\prime}$,
and $\lambda^{\prime\prime}$ respectively. Since the CKM elements
depend on the phase differences $\tilde{\Delta}$,
$\tilde{\Delta}^{\prime}$ and $\tilde{\Delta}^{\prime\prime}$
defined above, we have the following four sets of values:
1)$(1.47726,1.262553, 1.4592)$ giving $(7.962^\circ,
-181.124^\circ,-1.299^\circ)$,
2)$(1.477385,1.26252, 1.459415)$ giving $(6.508^\circ,
-180.92^\circ,-1.0631^\circ)$,
3)$(1.4775,1.2625, 1.45957)$ giving $(5.0566^\circ,
-180.715^\circ,-0.8269^\circ)$,
4)$(1.47753,1.262495, 1.45965)$ giving $(4.334^\circ,
-180.613^\circ,-0.7088^\circ)$,
where in (1) to (4) the first and second sets correspond
to $(\lambda, \lambda^{\prime},\lambda^{\prime\prime})$
and $(\tilde{\Delta},\tilde{\Delta}^{\prime},
\tilde{\Delta}^{\prime\prime})$ respectively. The results
are:1) $|V_{us}|=0.22$, $|V_{cb}|=0.0476$,
$|V_{ub}/V_{cb}|=0.0834$, $|V_{cs}|=0.9739$,
$|V_{cd}|=0.222$, $|V_{ud}|=0.9745$, $\hat{B}_K=0.555$,
2)$|V_{us}|=0.221$, $|V_{cb}|=0.0459$,
$|V_{ub}/V_{cb}|=0.0827$, $|V_{cs}|=0.9737$,
$|V_{cd}|=0.2228$, $|V_{ud}|=0.9743$, $\hat{B}_K=0.6385$,
3)$|V_{us}|=0.222$, $|V_{cb}|=0.0444$,
$|V_{ub}/V_{cb}|=0.082$, $|V_{cs}|=0.9736$,
$|V_{cd}|=0.2237$, $|V_{ud}|=0.9741$, $\hat{B}_K=0.794$,
4)$|V_{us}|=0.2225$, $|V_{cb}|=0.04378$,
$|V_{ub}/V_{cb}|=0.0816$, $|V_{cs}|=0.9736$,
$|V_{cd}|=0.224$, $|V_{ud}|=0.974$, $\hat{B}_K=0.913$,
for the four cases. In all four cases, the mass eigenvalues
are (at $1 GeV$) $m_U=1.28 TeV$, $m_t=258 GeV$, $m_c=
1.5 GeV$, $m_u=5.16 MeV$, $m_D=980 GeV$, $m_b=6.15 GeV$,
$m_s=147 MeV$, $m_d=8.6 MeV$. The values of $\hat{B}_K$
were obtained using the experimental value of $\epsilon$
namely 0.00227.
In this preliminary analysis
the physical top quark mass is $m^{p}_t\approx153 GeV$.
A better way to present the above results would be to use
a plot of the CKM elements versus $\hat{B}_K$ but the trend
can already be seen from the above numbers.
The latest lattice computation of the Isgur-Wise function \cite {isgur}
give $|V_{cb}|\sqrt{\tau_B/1.48ps}= 0.038^{+2+8}_{-2-3}$ \cite {ukqcd},
$|V_{cb}|\sqrt{\tau_B/1.53ps}= 0.044\pm0.005\pm0.007$ \cite {bernard},
and $\hat{B}_K=0.825\pm0.027\pm0.023$ using staggered fermions
\cite{sharpe}. Taken at face value,
$\hat{B}_K$ tends to favor $|V_{us}|\approx 0.222$,
$|V_{cb}|\approx 0.044$, $|V_{ub}/V_{cb}|\approx 0.082$,
$|V_{ud}|\approx 0.974$, $|V_{cs}|\approx 0.9736$,
$|V_{cd}|\approx 0.224$. Further experimental and theoretical
efforts are clearly needed to pinpoint these values.

One might also ask about the sensitivity of the top quark
mass prediction to the value of the fourth-generation mass.
Our preliminary results indicate that, for the fourth-generation
quark mass not too far from 1 TeV, the result is not very
sensitive to the precise value of that mass since having
fixed $\lambda_U$ and $\lambda_D$, one has $m_t \approx
m_c/\lambda_U^2$ and therefore $m_t$ is more sensitive
to $\lambda_U$ and $m_c$ than $m_U$. A more detailed investigation
of the dependence on $m_U$ for a larger range of values will
be carried out separately. The results presented here are
for $m_U$ around 1 TeV.

The lepton sector is identical in form to the quark sector
and the results can be obtained with
the replacement $h_{1,2}\rightarrow l_{1,2}$. A rough estimate
gives $\lambda_E\approx 1.4 \lambda_D (m_b^0/m_{\tau}^0)
(m_N^0/m_U^0)^3$ and $\lambda_N\approx 1.4 \lambda_E
(\lambda_U/\lambda_D)$. Here $m_N^0$ is the Dirac bare mass
of the fourth-generation neutrino.
Since $|V_{us}|\approx \lambda_U +
\lambda_D$, the leptonic version is $|V_{\nu_e\mu}|\approx
\lambda_N+ \lambda_E$. A back of the envelope estimate gives
$BR(K_L\rightarrow \mu e) \approx (\lambda_N+\lambda_E)^2
BR(K_L\rightarrow \mu \mu)$. With $BR(K_L\rightarrow \mu \mu)=
7.3\times10^{-8}$ and $BR(K_L\rightarrow \mu e)<9\times10^{-11}$,
we find $m_N^0< 451$ GeV. For the sake of estimate,the mass
eigenvalue is assumed to be similar to that obtained for $m_U$,
namely a fator of 1.2 of its input, giving $m_{Nphys}< 541$
GeV. Assuming a see-saw mechanism of the form $m_{\nu_4}=
m_{Nphys}^2/\cal{M}$ and requiring that $m_{\nu_4} > 46$ GeV,
one obtains $\cal{M} <$ 6.4 TeV. A more detailed calculation
would presumably gives a bound on the Majorana scale (at least
for the fourth generation) not too far
from the previous value. Also it is not hard to arrange the masses
of the other three neutrinos to be much lighter than the fourth one.
A host of interesting phenomena might
be studied such as an intriguing possibility of direct CP-violation
in $\tau$-decay, presumably in some kind of $\tau$ factory.

This work was supported in part by the U. S. Department of Energy
under Grant No. DE-A505-89ER40518.


%
%

%
%

\end{document}